\documentclass[12pt,thmsa]{article}
\usepackage{sw20elba}
\usepackage{amsmath}
\usepackage{amsfonts}
\usepackage{amssymb}
\usepackage{graphicx}

\begin{document}

\author{T. Tomita, J. J. Hamlin, and J. S. Schilling
\and \textit{Department of Physics, Washington University}
\and \textit{C. B. 1105, One Brookings Dr., St. Louis, MO 63130}\vspace*{0.6cm}
\and D. G. Hinks and J. D. Jorgensen
\and \textit{Materials Science Division, Argonne National Laboratory }
\and \textit{9700 South Cass Avenue, Argonne, IL 60439}\vspace*{0.6cm}}
\title{The Dependence of $T_{c}$ on Hydrostatic Pressure in Superconducting MgB$_{2} $}
\date{(submitted for publication on March 26, 2001)}
\maketitle
\begin{abstract}
The dependence of $T_{c}$ on hydrostatic (He-gas) pressure for superconducting
MgB$_{2}$ has been determined to 0.7 GPa. We find that $T_{c}$ decreases
linearly and reversibly under pressure at the rate $dT_{c}/dP\simeq
-1.11\pm0.02$ K/GPa. \ These studies were carried out on the same sample used
in earlier structural studies under He-gas pressure which yielded the bulk
modulus $B=147.2\pm0.7$ GPa. \ The value of the logarithmic volume derivative
of $T_{c}$ is thus accurately determined, $d\ln T_{c}/d\ln V=+4.16\pm0.08,$
allowing quantitative comparison with theory. \ The present results support
the emerging picture that MgB$_{2}$ is a BCS superconductor with
electron-phonon pairing interaction.
\end{abstract}

\newpage

The recent discovery\cite{r1} of superconductivity in MgB$_{2}$ at
$T_{c}\simeq$ 39 K has sparked worldwide a torrent of experimental and
theoretical activity, reminiscent of the frenzy following the observation
\cite{r2} of superconductivity in La-Ba-Cu-O at a comparable temperature
almost 15 years ago. \ Replacing $^{10}$B with $^{11}$B results in a sizeable
isotope shift\cite{r3} to lower temperatures which points to BCS
superconductivity. \ Other experiments, such as heat capacity\cite{r4,r4'},
photoemission spectroscopy\cite{r5}, and inelastic neutron
scattering\cite{r5',r6'} also support the picture that MgB$_{2}$ is a
phonon-mediated superconductor in the weak-to-moderate coupling regime.

High pressure studies traditionally play an important role in
superconductivity. \ A large magnitude of the pressure derivative $dT_{c}/dP$
is a good indication that higher values of $T_{c}$ may be obtained through
chemical means. \ It is not widely appreciated, however, that the pressure
dependence $T_{c}(P),$ like the isotope effect, contains valuable information
on the superconducting mechanism itself. \ For example, in simple-metal BCS
superconductors, like Al, In, Sn, and Pb, $T_{c}$ invariably decreases under
pressure due to the reduced electron-phonon coupling from lattice
stiffening\cite{r6}. \ More generally, an accurate determination of the
dependence of both $T_{c}\ $and the lattice parameters on pressure yields the
functional dependence $T_{c}=T_{c}\left[  a(P),b(P),c(P)\right]  \ $which
provides a critical test of theoretical models. \ Hirsch\cite{r7} and Hirsch
and Marsiglio\cite{r7'} have applied a theory of hole superconductivity to
MgB$_{2}$ and predicted that for an optimally doped sample $T_{c}$ should
increase with pressure, in contrast to the expected decrease in $T_{c}$ from
lattice stiffening.

Precise structural data on MgB$_{2}$ at room temperature (RT) have recently
been obtained by Jorgensen \textit{et al.}\cite{r10} for hydrostatic pressures
to 0.6 GPa in a He-gas neutron diffraction facility which yield the
anisotropic compressibilities $d\ln a/dP=-1.87\times10^{-3}$ GPa$^{-1}$, $d\ln
b/dP=-3.07\times10^{-3}$ GPa$^{-1}$, and the bulk modulus $B=147.2\pm0.7$ GPa;
the compressibility along the $c$ axis is thus significantly (64\%) larger
than that along the $a$ axis. \ The binding within the boron layers is
evidently much stronger than between the layers. \ These results are in
reasonable agreement with electronic structure calculations by Loa and
Syassen\cite{loa}.

Recent synchrotron x-ray diffraction studies at RT in a diamond-anvil-cell
(DAC) to much higher pressures (8 - 12 GPa) using dense He\cite{goncharov} or
methanol-ethanol\cite{r8} as hydrostatic pressure media yield the bulk moduli
$B=155\pm10$ GPa and $151\pm5$ GPa, respectively, in agreement, within
experimental error, with the He-gas study\cite{r10}. \ A further DAC
study\cite{r9} to 7 GPa with silicon oil as pressure medium gives the
significantly smaller value $B=120\pm5$ GPa, even though the reported 2.9\%
decrease in unit cell volume upon applying 6.5 GPa pressure is \textit{less}
than the 4\% decrease found in the other two DAC studies\cite{goncharov,r8}.
\ The strong deviation of the ``silicon oil'' data\cite{r9} may arise from
difficulties in extrapolating the data to zero pressure to obtain the bulk
modulus and/or from shear stresses arising from the solidification of the
silicon oil under pressure. \ The relatively large compressibility anisotropy
in MgB$_{2}$ mandates the use of hydrostatic pressure in quantitative studies,
since shear stresses applied by non-hydrostatic pressure media to an
elastically anisotropic sample can lead to erroneous results.

Several studies of the dependence of $T_{c}$ on pressure have appeared for
MgB$_{2}$. \ In the first experiment utilizing a fluid pressure medium
(Fluorinert), Lorenz \textit{et al.}\cite{r11} report that $T_{c}$ decreases
linearly with pressure to 1.8 GPa at the rate $dT_{c}/dP\simeq-1.6$ K/GPa.
\ On the other hand, using the same pressure medium, Saito \textit{et al.}
\cite{r12} find the more rapid decrease $dT_{c}/dP\simeq-2.0$ K/GPa. \ Both
groups cite their results to argue that MgB$_{2}$ is a BCS phonon-mediated
superconductor\cite{r11,r12}, as is also argued by Loa and Syassen\cite{loa}.
\ In further experiments to 25 GPa utilizing the solid pressure medium
steatite, Monteverde \textit{et al.}\cite{monte} find that $T_{c}$ decreases
under pressure at differing initial rates (-0.35 K/GPa to -0.8 K/GPa),
$T_{c}(P)$ showing a quadratic behavior for two of the four samples studied;
each of the four samples was prepared using a different synthesis procedure.

In the present experiment, $T_{c}(P)$ is determined to 0.7 GPa in an ac
susceptibility measurement using a He-gas pressure system; the helium pressure
medium remains fluid at $T_{c}\simeq$ 39 K up to 0.5 GPa and thus applies true
hydrostatic pressure (no shear stresses) to the sample. \ We find that $T_{c}$
decreases linearly and reversibly with pressure at the rate $dT_{c}%
/dP\simeq-1.11\pm0.02$ K/GPa. \ Implications for the nature of the
superconducting state are discussed.

The powder sample for this study was taken from the same mother sample used in
parallel neutron diffraction studies\cite{r10}. \ It is made using
isotopically-enriched $^{11}$B (Eagle Picher, 98.46 atomic \% enrichment). \ A
mixture of $^{11}$B powder (less than 200 mesh particle size) and chunks of Mg
metal were reacted for 1.5 hours in a capped BN crucible at 800$^{\circ}$C
under an argon atmosphere of 50 bar. \ As seen in Fig. 1, the resulting sample
displays sharp superconducting transitions in the ac susceptibility with full
shielding and an onset temperature at ambient pressure $T_{c}(0)\simeq$ 39.25
K. \ Both x-ray and neutron diffraction data show the sample to be single
phase with the AlB$_{2}$-type structure.

The present high pressure studies were carried out using a He-gas
high-pressure system (Harwood) to 1.4 GPa; the pressure is determined by a
calibrated manganin gauge in the compressor system at ambient temperature.
\ The superconducting transition of the 8.12 mg MgB$_{2}$ powder sample is
measured by the ac susceptibility technique using a miniature
primary/secondary coil system located inside the 7 mm I.D. bore of the
pressure cell. \ A small Pb sphere with 1.76 mm dia. (38.58 mg) is also
inserted in the coil system for susceptibility calibration purposes; for
selected data the superconducting transition temperature of this Pb sphere
serves as an internal manometer\cite{r20} to check the pressure indicated by
the manganin gauge. \ The CuBe pressure cell (Unipress), which is connected to
the compressor system by a 3 mm O.D. $\times$ 0.3 mm I.D. CuBe capillary tube,
is inserted into a two-stage closed-cycle refrigerator (Leybold) operating in
the temperature range 2 - 320 K. \ The pressure can be changed at any
temperature above the melting curve $T_{m}$ of\ the helium pressure medium
(for example, $T_{m}\simeq$ 13.6 K at 0.1 GPa and $T_{m}\simeq$ 38.6 K at 0.50
GPa\cite{r14}). \ For pressures above 0.5 GPa, $T_{m}$ lies above $T_{c}$; the
slight pressure drop (few 0.01 GPa's) on cooling from $T_{m}$ to $T_{c}$ is
estimated using the isochores of He\cite{r14}. \ All pressures are determined
at the temperature $T_{c}\simeq$ 39 K. \ Further details of the experimental
techniques are given elsewhere\cite{r15}.

In Fig. 1 are shown representative examples of the superconducting transition
in the ac susceptibility at both ambient and high pressure. \ With increasing
pressure the narrow transition is seen to shift bodily to lower temperatures,
allowing a determination of the pressure-induced shift in $T_{c}$ to within
$\pm$ 10 mK. \ Remarkably, close inspection of the data for 0.50 GPa reveals a
slight shift in the transition curve near its midpoint, accurately marking the
position of the melting curve of helium ($T_{m}\simeq$ 38.6 K) at this pressure.

In Fig. 2, $T_{c}$ is plotted versus applied pressure to 0.7 GPa and is seen
to follow a highly linear dependence $dT_{c}/dP\simeq-1.11\pm0.02$ K/GPa.
\ The first data point 1$^{\prime}$ was obtained after first applying a
pressure of $\sim$ 0.7 GPa at RT before cooling down to low temperatures
($\sim$ 60 K) and reducing the pressure, yielding $T_{c}\simeq$ 38.88 K at
0.341 GPa. Point 2$^{\prime}$ was measured after releasing the pressure at low
temperature, giving $T_{c}\simeq$ 39.25 K at ambient pressure (0 GPa); no
change in $T_{c}$ occurred after intermittently warming the sample to RT
(point 3). \ Further data were obtained following pressure changes at both RT
(unprimed data) and low temperature (primed data). \ As is observed for the
vast majority of superconducting materials without phase change, the
dependence of $T_{c}$ on pressure for MgB$_{2}$ is single-valued and does not
depend on the pressure/temperature history of the sample; such history effects
do occur in certain high-$T_{c}$ oxides containing mobile species at
RT\cite{relaxation}. \ We thus find that for He-gas pressure changes at both
ambient and low temperature, $T_{c}(P)$ for MgB$_{2}$ is a linear, reversible
function of pressure to 0.7 GPa \cite{note1}.

In the present experiments the sample is surrounded by fluid helium near
$T_{c}\simeq$ 39 K for all data taken at pressures $P\leq0.50$ GPa so that the
slope $dT_{c}/dP\simeq$ -1.11 K/GPa gives the true hydrostatic pressure
dependence of $T_{c}$. \ The fact that the sample is in solid helium for
$P>0.5$ GPa is seen in Fig. 2 to have no effect on the pressure dependence
$T_{c}(P);$ indeed, solid helium is the softest solid known. \ Our value of
$dT_{c}/dP$ differs significantly from those of other
groups\cite{r11,r12,monte} (see discussion above) using pressure media which
are either solid at RT or freeze upon cooling down at temperatures well above
$T_{c}.$

In view of the strong compressibility anisotropy\cite{r10} and the sizeable
anharmonicity and non-linear electron-phonon coupling\cite{r6'} anticipated
for MgB$_{2}$, it is likely that shear stresses of sufficient magnitude will
cause appreciable changes in the pressure dependence of $T_{c},$ as observed
for other anisotropic substances such as the superconducting
oxides\cite{stefan} and organic superconductors\cite{r16}. \ The largest shear
stresses are generated by changing the pressure on a solid pressure medium,
such as steatite, or using no pressure medium at all. \ The shear stresses
generated in cooling Fluorinert or other comparable liquids through the
melting curve are admittedly much smaller and depend on details of the
individual experimental procedures used, such as the cooling rate, change in
applied force upon cooling, etc. \ Only experiment can determine whether or
not the $T_{c}(P)$ dependences measured in the Fluorinert
experiments\cite{r11,r12} are influenced by shear stresses. \ To exclude the
possibility of sample-dependent effects, such experiments should be carried
out on a single sample. \ Lorenz \textit{et al.}\cite{lorenz2} have very
recently carried out He-gas high-pressure studies on the same sample studied
by them previously in Fluorinert\cite{r11} and find a value of $dT_{c}/dP$
equal to their previous value (-1.6 K/GPa), within experimental error; further
measurements in the same He-gas system on a second, high-quality sample
yielded the dependence $dT_{c}/dP\simeq-1.07$ K/GPa, a value very close to our
present result. \ This finding lends support to the observation by Monteverde
\textit{et al.} \cite{monte} that $dT_{c}/dP$ in MgB$_{2}$ may be sample dependent.

The present $T_{c}(P)$ studies and parallel high-pressure structural studies
by Jorgensen \textit{et al.}\cite{r10} were both carried out on the same
MgB$_{2}$ sample over the same He-gas pressure range, thus allowing an
accurate determination of the change in $T_{c}$ with lattice parameter. \ The
change in $T_{c}$ with unit cell volume, for example, is given by
\begin{equation}
\frac{d\ln T_{c}}{d\ln V}=\frac{B}{T_{c}}\left(  \frac{dT_{c}}{dP}\right)
=+4.16\pm0.08,
\end{equation}
using the above values $dT_{c}/dP\simeq-1.11\pm0.02$ K/GPa, $B=147.2\pm0.7 $
GPa, and $T_{c}=39.25$ K.

We will now discuss the implications of this result for the nature of the
superconducting state in MgB$_{2}.$ \ First consider the McMillan
equation\cite{r17} \newline $T_{c}\simeq(\left\langle \omega\right\rangle
/1.20)\exp\left\{  \left[  -1.04(1+\lambda)\right]  /\left[  \lambda-\mu
^{\ast}(1+0.62\lambda)\right]  \right\}  ,$ valid for strong coupling
($\lambda\lesssim1.5),$ which connects the value of $T_{c}$ with the
electron-phonon coupling parameter $\lambda,$ an average phonon frequency
$\left\langle \omega\right\rangle ,$ and the Coulomb repulsion $\mu^{\ast},$
which we assume to be pressure independent. \ Taking the logarithmic volume
derivative of $T_{c}$, we obtain the simple relation
\begin{equation}
\frac{d\ln T_{c}}{d\ln V}=-\gamma+\Delta\left\{  \frac{d\ln\eta}{d\ln
V}+2\gamma\right\}  ,
\end{equation}
where $\gamma\equiv-d\ln\left\langle \omega\right\rangle /d\ln V$ is the
Gr\"{u}neisen parameter, $\eta\equiv N(E_{f})\left\langle I^{2}\right\rangle $
is the Hopfield parameter\cite{r17''} given by the product of the electronic
density of states and the average squared electronic matrix element, and
$\Delta=1.04\lambda\left[  1+0.38\mu^{\ast}\right]  /\left[  \lambda-\mu
^{\ast}(1+0.62\lambda)\right]  ^{2}.$

Eq. (2) has a simple interpretation. \ The first term on the right, which
comes from the prefactor to the exponent in the McMillan expression for
$T_{c}$, is usually small relative to the second term, as will be demonstrated
below. \ The sign of the logarithmic derivative $d\ln T_{c}/d\ln V$,
therefore, is determined by the relative magnitude of the two terms in the
curly brackets. \ The first ``electronic'' derivative is negative ($d\ln
\eta/d\ln V\approx-1$ for simple metals (s,p electrons) \cite{r17'}, but
equals -3 to -4 for transition metals (d electrons)\cite{r17''}), whereas the
second ``lattice'' term is positive (typically $2\gamma\approx3-5).$ \ Since
in simple-metal superconductors, like Al, In, Sn, and Pb, the lattice term
dominates over the electronic term, and\ $\Delta$ is always positive, the sign
of $d\ln T_{c}/d\ln V$ is the same as that in the curly brackets, namely
positive; this accounts for the universal decrease of $T_{c}$ with pressure
due to lattice stiffening in simple metals. \ In selected transition metals
the electronic term may become larger than the lattice term, in which case
$T_{c}$ would be expected to increase with pressure, as observed in
experiment\cite{r17''}.

Let us now apply Eq. (2) in more detail to a canonical BCS simple-metal
superconductor. $\ $In Sn, for example, $T_{c}$ decreases under pressure at
the rate $dT_{c}/dP\simeq$ -0.482 K/GPa which leads to $d\ln T_{c}/d\ln
V\simeq+7.2$ \cite{r20}. \ Inserting $T_{c}(0)\simeq$ 3.73 K, $\left\langle
\omega\right\rangle \simeq110$ K \cite{r21}, and $\mu^{\ast}=0.1$ into the
above McMillan equation, we obtain $\lambda\simeq0.69$ from which follows that
$\Delta\simeq2.47.$ \ Inserting these values into Eq. (2) and setting
$d\ln\eta/d\ln V\approx-1$ from above for simple metals, we can solve Eq. (2)
for the Gr\"{u}neisen parameter to obtain $\gamma\simeq+2.46,$ in reasonable
agreement with experiment for Sn ($\gamma\approx+2.1)$\cite{r20}. \ Similar
results are obtained for other simple-metal superconductors.

We now repeat the same calculation with the McMillan equation for MgB$_{2}$
using the logarithmically averaged phonon energy from inelastic neutron
studies\cite{r5'} $\left\langle \omega\right\rangle =670$ K, $T_{c}%
(0)\simeq39.25$ K, and $\mu^{\ast}=0.1,$ yielding $\lambda\simeq0.90$ and
$\Delta\simeq1.75$. \ From Eq. (1) we have $d\ln T_{c}/d\ln V=+4.16.$ \ Since
the pairing electrons in MgB$_{2}$ are believed to be s,p in
character\cite{r22}, we set $d\ln\eta/d\ln V\approx-1$ \cite{note10}.
\ Inserting these values into Eq. (2), we find $\gamma\simeq2.36,$ in
reasonable agreement with the value $\gamma\approx2.9$ from Raman spectroscopy
studies\cite{goncharov} or $\gamma\approx2.3$ from \textit{ab initio
}electronic structure calculations on MgB$_{2}$ \cite{r18}. \ If, on the other
hand, one were to use the same bulk modulus but the pressure derivative
$dT_{c}/dP\simeq-2.0$ K/GPa from Saito \textit{et al.}\cite{r12}, one obtains
from Eq. (2) the unusually high value $\gamma\simeq3.7.$

In extensive specific heat\cite{r4'} and high-resolution photoemission
studies\cite{tsuda} on MgB$_{2},$ evidence is found for a multicomponent
superconducting gap; the latter study also reports an inconsistency in the
values of the electron-phonon coupling constant from McMillan's equation and
the renormalization of the electronic density-of-states. \ These results call
into question the suitability of the isotropic McMillan equation for
describing this system.

Under the above assumptions, we thus conclude that the rate of decrease of
$T_{c}$ with pressure found in the present experiments on MgB$_{2}$ is
consistent with BCS phonon-mediated superconductivity. \ The authors hope that
the accurate determination of the volume dependence of $T_{c}$ in this work,
$d\ln T_{c}/d\ln V=+4.16\pm0.08,$ will stimulate \textit{ab initio}
theoretical calculations. \ At first glance the present results appear to be
inconsistent with the model of Hirsch and Marsiglio\cite{r7,r7'} which
predicts that $T_{c}$ should increase with pressure for optimally doped
samples. \ However, within their model, $T_{c}$ for a non optimally doped
sample may decrease if sufficient change in the carrier concentration occurs
when pressure is applied. \ Further studies, such as high-pressure Hall effect
measurements, are necessary to clarify this possibility.\vspace{0.4cm}

\noindent\textbf{Acknowledgments}

Work at Washington University supported by NSF grant DMR 98-03820 and that at
the Argonne National Laboratory by the U.S. Department of Energy, Office of
Science, contract No. W-31-109-ENG-38.

\newpage

\section{\textbf{Figure Captions}}

\bigskip\ 

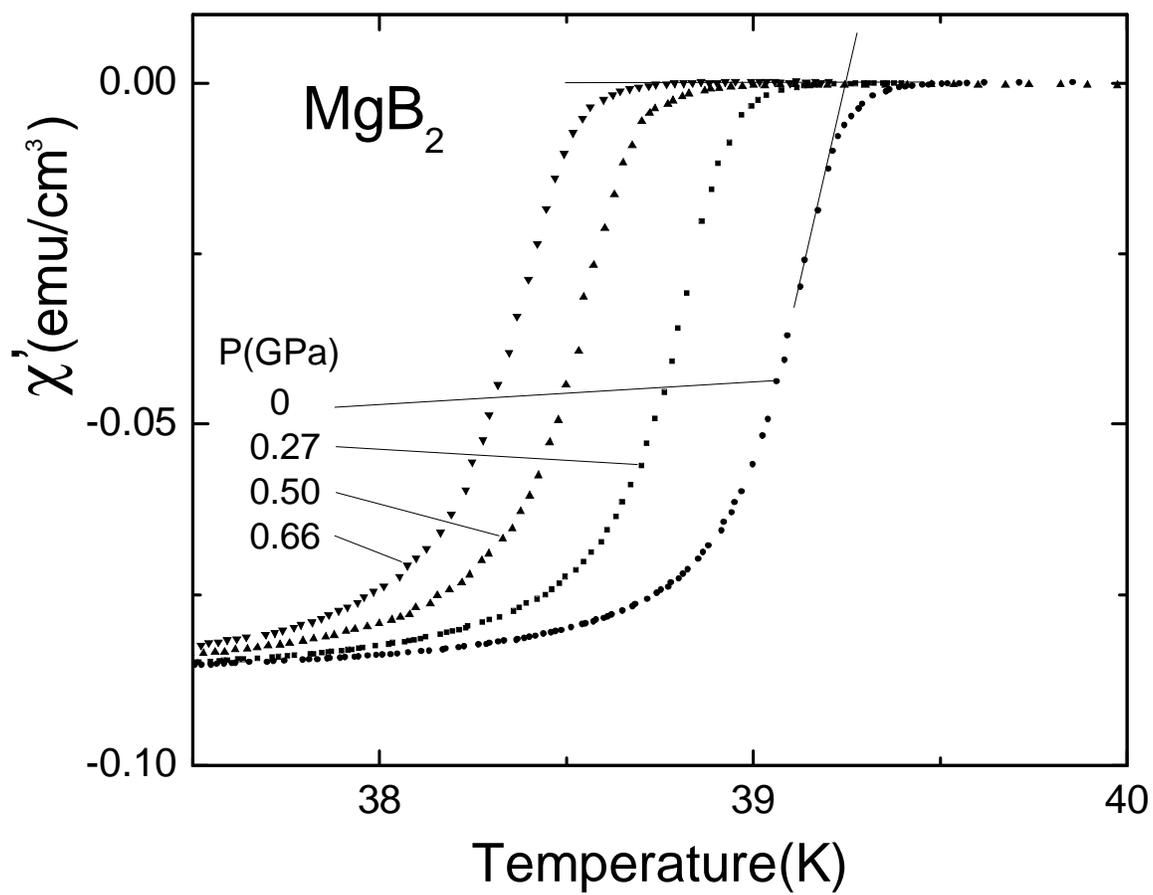
\noindent\textbf{Fig. 1. \ }Real part of the ac susceptibility of MgB$_{2}$
versus temperature at ambient and high pressures. \ The applied magnetic field
is 0.113 Oe (rms) at 1,023 Hz. \ Intercept of straight tangent lines defines
superconducting onset at ambient pressure $T_{c}(0)\simeq$ 39.25 K. \ No
correction is made for demagnetization effects.\bigskip

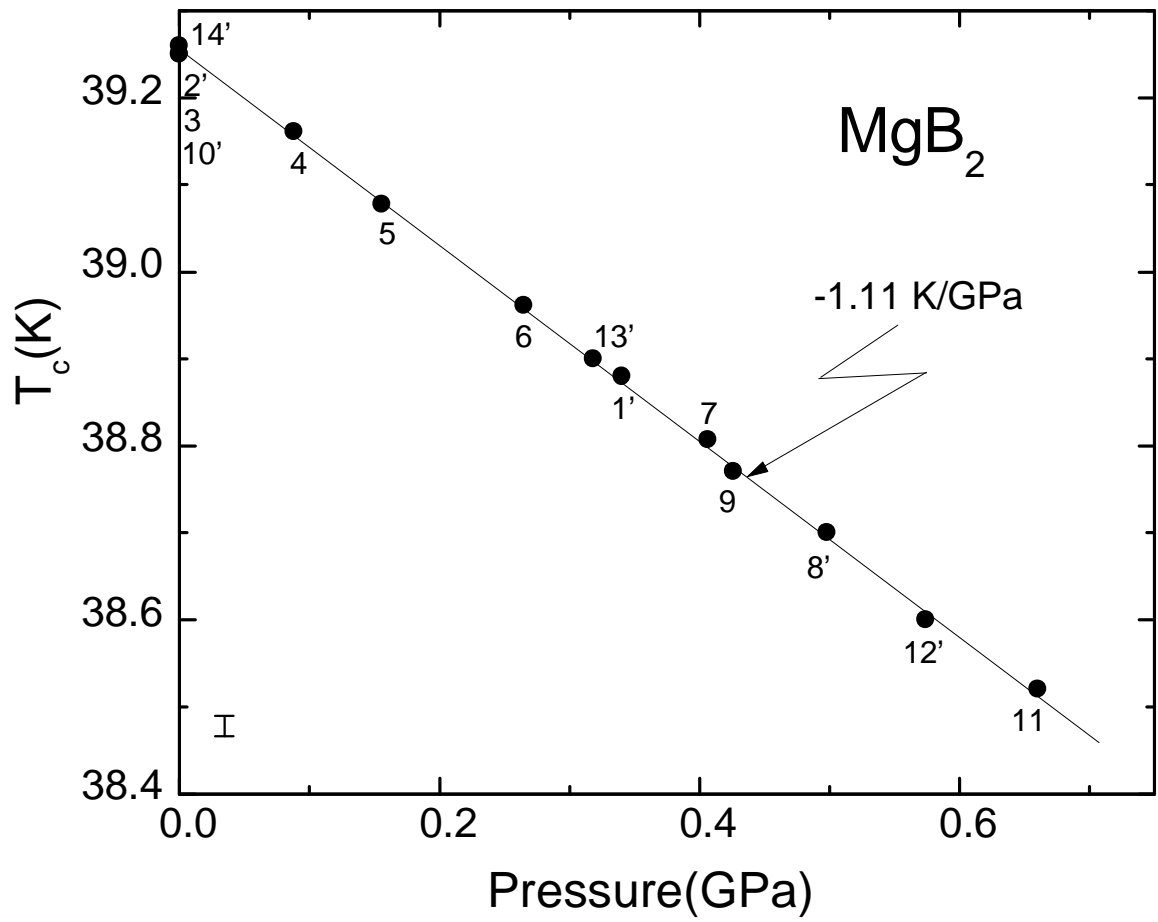
\noindent\textbf{Fig. 2. \ }Superconducting transition temperature versus
applied pressure. \ Numbers give order of measurement. \ Data for pts.
2$^{\prime}$, 6, 8$^{\prime}$, and 11 are shown in Fig. 1. \ A typical error
bar for $T_{c}$ ($\pm0.01$ K) is given in lower left corner; the error in
pressure is less than the symbol size. \ Pressure was either changed at RT
(unprimed numbers) or at low temperatures $\sim$ 60 K (primed numbers).\newpage%


\begin{thebibliography}{99}
\bibitem {note1}In a separate experiment on MgB$_{2}$ to 0.077 GPa pressure in
Neon gas, an identical $T_{c}(P)$ dependence was obtained.
\end{thebibliography}
\end{document}